\documentclass{article}

\usepackage{PRIMEarxiv}

\usepackage[utf8]{inputenc} 
\usepackage[T1]{fontenc}    
\usepackage{hyperref}       
\usepackage{url}            
\usepackage{booktabs}       
\usepackage{amsfonts}       
\usepackage{nicefrac}       
\usepackage{microtype}      
\usepackage{fancyhdr}       
\usepackage{graphicx}       
\graphicspath{{media/}}     

\title{Which tweets `deserve' to be included in news stories?\\Chronemics of tweet embedding
 }

\author{Munif Ishad Mujib \\
 Drexel University \\
 {\underline{mim52@drexel.edu}} \\\And
 Asta Zelenkauskaite \\
 Drexel University \\
 Vilnius Gediminas Technical University \\
 {\underline{az358@drexel.edu} }\\\And 
 Jake Ryland Williams \\
 Drexel University \\
 {\underline{jw3477@drexel.edu}} \\}
\begin{document}
\maketitle

\begin{abstract}
The use and selection of user-generated social media content, specifically tweets, as a news source has become an integral part of news production practice. Yet, the mapping and the extent of the nature of the practices in which news outlets integrate social media use is still lacking. This study focuses on the pressures of immediacy on the media ecosystems, i.e., as organizational practices of news outlets that make choices related to social media content integration. By analyzing a large corpora of news outlets that have embedded tweets, this study analyzes tweet embedding practices by specifically focusing on the concept of chronemics, conceptualized here as the time needed to embed tweets. Temporal constraints are particularly pressing for journalistic practices, given the continuous pressures of 24/7 news cycle. We ask two main questions: which types of outlets are quicker to embed tweets, and which types of users' tweets are more likely to be embedded quickly?
\end{abstract}

\section{Introduction}
The digitization of news outlets has had numerous implications for news production: on one hand, content can be continuously updated (\cite{kutz2005microlongitudinal}); on the other hand, it has opened up new opportunities to embed external digital sources into news stories. Direct inclusion of external sources such as social media (e.g. tweet embedding) is a growing practice in news organizations (\cite{oschatz2021twitter}). This study investigates new conditions of journalism today explicated by scholars like \cite{seuri2022happens} as platformization and the blurring of journalism’s boundaries. It is based on the premise that the inclusion of user-generated social media content in mass media creates new ecosystems that are in flux, as argued by \cite{zelenkauskaite2018value}. 
Among various social networks that are the sources of user-generated content embedded in news, Twitter holds particular significance for journalists, not only through the use of Twitter to disseminate information (\cite{enli2018social}), but also by using tweets as a type of evidence for a particular story (\cite{broersma2012social}). 
In this work, the process and pace of tweet integration in news stories is treated through an ecosystem-as-structure approach that involves actors, technology, and value creation through publication at the right time, as in \cite{adner2017ecosystem}. The focus here is to analyze a specific practice---tweet selection for news---elucidating how journalists select from the constant stream of content being generated by users and embed them as interactive data objects within web versions of news articles. 
In order to investigate the tweet integration phenomenon at a large scale, we designed and implemented a streaming data acquisition system that collects digital news articles and embedded social media content, particularly content from Twitter, discovered within those articles. We call this acquisition system \emph{NewsTweet}. We present a study of the tweet embedding phenomenon performed by analyzing approximately two years' worth of data collected by NewsTweet from 2019-2021 (\cite{mujib2020newstweet}).

\subsection{Chronemics in online spaces}

This study's primary focus is on the temporal dimension, i.e., the speed with which tweets are incorporated in news stories. This is theoretically conceptualized through chronemics in online spaces, defined by the expectations of immediacy (\cite{kalman2005email}). The concept of ``speed'' of tweet integration arises from assumptions based on the concept of chronemics, which has been extensively analyzed in studies of computer-mediated communication (\cite{kalman2005email}). We suggest that chronemics in online spaces matter because it influences the pace in which content is distributed and received. Chronemics assumes that there is a certain temporal ``threshold'' that makes the content, i.e., a tweet, temporally relevant for the recipients of that content (\cite{kalman2005email}). In this study, we ask an overarching question: how do chronemics play out in social media, i.e., tweet integration as a source in a given news story? Based on the existing literature on the topic, there are three approaches through which we considered chronemics to have an effect on tweet selection for the news story in this study: 1) the news outlet type; 2) the news outlet overall news posting pacing (we conceptualized them as frequent or infrequent); and 3) the author of the tweet. 



Assumptions for this study are that the news outlets as media ecosystems have different capacities (due to rate of news story creation by the outlet and type), thus resulting in different pace or speeds between tweet posting and their integration in a given news story. We operate on two assumptions. The first one relates to the recency of the tweet as a source integration hypothesis--where more recent sources are more valued by news organizations as a source. The second is the democratizing assumption of twitter for news: theoretically, we can assume that due to egalitarian access to Twitter, anyone who posts on Twitter has the same right to be selected, depending on the content relevance. Alternative to this premise is the assumption of "trustworthiness" where traditional sourcing,  trusting the source (e.g., because of its institutional prominence or celebrity status as argued in \cite{berkowitz2019reporters}) will be observed in the account choices for tweets. This assumes that journalists ascribe varying levels of immediacy to users' tweets depending on their authorship, e.g., institutional standing and journalistic affiliation. We assume that news outlets treat trustworthiness differently for users, here referred to as authors of a given tweet, or, the \textit{stakeholders}. 

Considering the tweet author as a stakeholder, we should likely expect that there are some who are more ``trusted'' by news organizations. Therefore, we hypothesize the following: Tweets that are embedded faster are from more ``trusted'' sources. Another assumption that drives this work is the outlet's capacity to produce news stories backed by sources. Thus, we assume that outlets that produce more stories will also be using tweets for sourcing, given the accessibility of tweets.  

\section{The mechanisms of socially-connected news production}
This section contextualizes journalistic chronemics practices by describing how Twitter is embraced in journalism. In particular we study Twitter as ``the \emph{de facto} tool of newsgathering'' (\cite{liu2016reuters}), providing sources akin to the live eye-witness reporting typical within journalism. Via this functionality---re-inscribed in social media platforms (\cite{zelenkauskaite2022facebook})---Twitter allows for real-time public sourcing of raw news material (\cite{newman2009rise,broersma2012social,lasorsa2012normalizing,hermida2013hashtagjournalism}). Similarly, sociotechnical configurations of Twitter such as the ``retweet'' function, which allows one user to share another's content with their followers, mimic traditional mass media ``broadcasting'' (\cite{ahmad2010twitter}). We study amalgamated tweet-broadcasts that are integrated in mass media. 

\subsection{Journalistic practices in using Twitter as a news source}
The democratizing hypothesis of twitter starts with the premise that tweets as an acceptable source have become commonplace as a part of established journalistic practices. By 2010, well-regarded and established print outlets such as \emph{The New York Times} and \emph{The Washington Post}, along with all major national television networks, were already using Twitter content as primary sources in some stories (\cite{moon2014routinizing}). 
Tweets have been used to fulfill the needs of real-time issues, e.g., reporting breaking news events. Examples of major news events where Twitter has been instrumental as a source medium in reporting are the 2010 Haiti earthquake, the 2011 Arab Spring uprisings, and the Black Lives Matter movement (\cite{newman2009rise,bruno2011tweet,hermida2012tweets,freelon2018black}). While it is also noted that tweets as a source were more frequently used when reporting focused on “less-serious” topics, such as art, sports, entertainment, and lifestyle, tweets produced by former U.S. President Donald Trump (until his eventual ban from the platform) were frequently reported on as well, often becoming news events unto themselves (\cite{mitchell2017covering}).

Similarly, differences in tweet use for news were found between traditional print-oriented outlets and web-only outlets (\cite{bane2019tweeting}), finding that web-only outlets are more likely to use tweets as sources of information rather than opinion or reaction. Thus, in this study we take into account the type of the news outlet and hypothesize that web-only outlets, due to multimodality of the content, will incorporate more tweets as a source regardless of the content.

Case studies on political reporting have found that in addition to providing \emph{vox populi} commentary, tweets were also being used as sources to shed light on the personal lives of public figures such as politicians and celebrities (\cite{broersma2013twitter}). 
One of the positive aspects of tweets as a source is that Twitter has allowed journalists to diversify the range of people and organizations that can be featured as sources, identifying a lowered barrier to entry for who can qualify as supplying evidence for a news story, thus, cited as a democratizing effect (\cite{paulussen2014social}). Thus, we hypothesize that there will be a range of users whose tweets will be included regardless of their "institutional backing", i.e., tweets' sourcing will not be gathered from just a few accounts, if tweet embeddings are treated as a democratizing instrument for journalism.

Due to velocity of upload and the recency aspect of the sourcing capacity of the ``crowds'', the use of Twitter content for breaking news has also been celebrated. However, this is often taken with a grain of salt for how readily available content impacts the journalistic principle of source verification (\cite{bruno2011tweet,vis2013twitter,thurman2013live,lecheler2016reevaluating}). The pressures of the 24-hour news cycle can lead to difficult choices for journalists in deciding between ``being fast'' and ``being right'' (\cite{gowing2009skyful,hermida2013hashtagjournalism}) or the tension between verification and publication (\cite{hermida2012tweets}). 

Given the temporal constraints and need to provide trusted sources discussed above, how do journalists select tweets as sources for news? Various technologies and automatic systems have been developed over the years in order to help journalists monitor and report on breaking news events (\cite{liu2016reuters,zubiaga2019mining,phuvipadawat2010breaking,schulz2013see,weiler2013event,wei2017finding,xie2016topicsketch}). Journalistic emergent practices for newsworthy tweets selection include the use of a combination of Twitter-provided and third-party tools to monitor the platform for news events and developments, going far beyond ``following'' other journalists and potentially newsworthy users. Yet, in this study if we start with the democratizing hypothesis of tweets in journalism, embedding speeds will not vary by author of tweet. And as a result, all users will be considered equally newsworthy to be embedded in the news story.

\subsection{Embedding tweets under time pressure}
In this study, we investigate whether there are differences in how outlets negotiate the immediacy considerations of tweet embedding given their orientations in terms of distribution and content focus. We hypothesize that there is indeed a race among outlets to embed newsworthy tweets. Our study investigates whether these differences in orientation translate to outlets ``winning'' the race---embedding tweets fastest---more often.

The guiding assumption is that newsrooms today are under immense time pressure to publish articles and capture web traffic, given the high value of immediacy (\cite{domingo2008interactivity}). Fast-paced sourcing is critical for the internet audience-driven immediacy expectations becoming dominant in the traditional newsroom cultures (\cite{usher2014making}). At times, many viral stories, or, ``low-hanging fruit'' (\cite{davies2011flat,saridou2017churnalism}), are produced by citing, reproducing, or otherwise referring to social media content. Journalism scholars also warned that multiple points of view in sourcing is the highest standard of reporting, but it has been found that this is not always achieved: one study found that single-source stories amounted to 76 percent of the analyzed story sample (\cite{o2008passive}).

Accessibility of social media content makes tweets into convenient sources. When news outlets report on events unfolding in real time, the overwhelming tendency is to publish and disseminate succinct stories first. However, news organizations are diverse in size and capacity, and hence ability for processing ever changing cycles of news. For example, in the early age of internet-enabled journalism, conventional outlets displayed a tendency to update websites with new stories at regular, slow intervals (such as once a day), giving rise to the publication of ``shovelware'' (\cite{scott2005contemporary}). This model of online publication tended to result in online news lagging behind traditional channels. As the online news ecosystem has evolved, growing to include discussion and engagement from users through comments, immediacy has become an expectation in online news (\cite{karlsson2011immediacy}). Thus, we expect variation between speed of tweet integration by outlet type.


\subsection{Filtering in tweet selection: speed and authors}
Story noteworthiness in journalism has been historically conceptualized as being achieved through the process of as gatekeeping as a key practice by the editor-in-chief who selects or discards stories to be published (\cite{white1950gate}). Journalistic gatekeeping has changed over the years, involving more of the constituents to make decisions about story filtering (\cite{shoemaker2009gatekeeping}). However, the notion of gatekeeping as selecting or filtering stories has been expanded and treated as a complex phenomenon that can be analyzed from various perspectives. From an institutional perspective, this affects the determination of newsworthiness for news stories as well as sources and event selection (\cite{shoemaker1996mediating,shoemaker2009gatekeeping}). Thus, the gatekeeping framework allows us to view the selection process as a wide or a narrow filter. While this process arises out of core journalistic values such as trustworthiness and the quality of information, the effect of gatekeeping is that in most cases, journalists rely on established sources, which in traditional journalism are institutional sources that are seen as credible by default (\cite{berkowitz2019reporters}).

Twitter grants journalists access to a much wider range of potential sources beyond the traditional, yet, journalists are found to adhere to many of the gatekeeping tendencies that formed in an earlier age of mass media, e.g., print media such as newspapers, even when using tweets as sources, preferring to cite public figures and prominent organizations that are thus vetted by their mere reputation as being credible (\cite{moon2014routinizing}). However, in specific types of news categories (often time-sensitive ones), some relaxation of these tendencies has been observed, leading to some authors suggesting that a reconceptualization of journalism is slowly taking place in that Twitter, for example, has been established as an online space where journalists gather the news and find sources, and then report the news and drive traffic to websites (\cite{hermida2012tweets}).

The uniquely public nature of discourse and information sharing on Twitter leads us to consider it in terms of the Habermasian principles of the public sphere, in particular,  of disregard for status and inclusivity (\cite{habermas1962structural}). While it is often claimed that Twitter has democratized information and opinion sharing by constructing a ``town square'' on the internet, in the context of newsworthiness, users face considerable gatekeeping from journalists. Source selection from Twitter by journalists has been demonstrated to favor ``elite'' voices (\cite{von2018sourcing}). The selection process for tweet embedding operates on the axes of both institutional sourcing credibility and ``specialty'' of content. A particular dynamic in social embedding is the reproduction of content from established journalists, giving rise to a community of journalists founded on credibility that is drawn from when tweets are chosen to be embedded. 

\section{Data Collection and Preprocessing}
NewsTweet is an ongoing collection of digital news articles. Articles listed on eight topical feeds in the Google News aggregator (Business, Entertainment, Health, Nation, Technology, Sports, World, and Headlines) are automatically collected, and embedded social media content is extracted. A total of 1,635,684 articles were collected over the collection period of this study (May 2019–June 2021). Embedded social content was extracted from these articles through automatic identification of specific HTML tags. This method of extraction ignored any unstructured references to social content such as through quotes in the article text. Embedded content volumes by platform are presented in table~\ref{tab:embeds_by_platform}. Tweet embeds made up nearly two-thirds of all embeds, and the volume of embedded tweets was more than double the next-most common embed type (YouTube videos). This firmly establishes Twitter's prominence in news over other social platforms. Overall, in the full NewsTweet dataset, 355,755 tweet embeddings were found, appearing across 149,901 articles representing approximately 9\% of all articles. These tweets were extracted from the articles, and extensive tweet metadata was collected using the Twitter API. Each article contained a publication timestamp provided by Google News. The accuracy of these timestamps was examined by sampling 160 random articles, and it was found that the timestamps were reliable in 91\% of the sampled articles. A set of outlets were identified as sources of unreliably-timestamped articles and discarded; this set consisted of a number of smaller international outlets as well as local television and radio station websites. Articles from these discarded outlets consisted of less than 1\% of the total collection. 

\begin{table}[h!]
    \caption{Volumes of embedded content by platform.}
    \small
    \renewcommand{\arraystretch}{1.4}
    \begin{center}
    \begin{tabular}{|c|r|r|}
    \hline
    \multicolumn{1}{|c|}{{\bf Platform}} & \multicolumn{1}{c|}{{\centering \bf Articles}} & \multicolumn{1}{c|}{{\bf Embeddings}}\\\hline
    Twitter & 149,901 (52.9\%) & 355,755 (64.4\%)\\
    YouTube & 95,059 (33.5\%) & 135,723 (24.6\%)\\
    Instagram & 34,525 (12.2\%) & 55,815 (10.1\%)\\
    Facebook & 1,911 (0.7\%) & 2,258 (0.4\%)\\
    Reddit & 1,177 (0.4\%) & 1,264 (0.3\%)\\
    TikTok & 766 (0.3\%) & 1,640 (0.2\%)\\\hline
    \bf{Total} & 283,339 (100.0\%) & 552,455 (100.0\%)\\\hline 
    \end{tabular}
    \end{center}
   \label{tab:embeds_by_platform}
\end{table}

\section{Methods}
Our study on the chronemics of tweet embedding was operationalized by the speed of news outlets to include a given tweet. For each embedded tweet, the time between tweet creation and article publication (i.e., embedding of the tweet) was computed and termed the \emph{embed delay}. The median embed delay was approximately 6 hours. The distribution appeared predictably heavy-tailed, with 77\% of embeddings occurring within 24 hours of tweet publication. It is worth pointing out that 3.5\% of embeds were found with ``negative'' embed delays, i.e., tweets were found embedded in articles that were published \emph{before} the tweets' creation times. Given that we treated these embeddings with negative embed times functioning with a potentially different institutional logic or practices, we did not include these in subsequent analyses, and they present an interesting avenue for future study.

\begin{table}[h!]
    \small
    \caption{Top embedding outlets overall.}
    \begin{center}
        \resizebox{\linewidth}{!}{
        \renewcommand{\arraystretch}{1.3}
        \begin{tabular}{|c|c|c|c|c|c|}
            \hline
            \textbf{Outlet} & \textbf{Distribution type} & \textbf{Content focus type} & \textbf{Articles with embeds} & \textbf{Median embed delay (hours)} \\ \hline
            yahoo.com & Online-only & General & 6,168 & 5.77 \\
            thehill.com & Online-only & General & 5,859 & 3.13 \\
            cbssports.com & Traditional (TV) & Sports & 5,550 & 1.25 \\
            usatoday.com & Traditional (Print) & General & 4,895 & 3.00 \\
                cnn.com & Traditional (TV) & General & 3,623 & 14.68 \\
            nbcsports.com & Traditional (TV) & Sports & 3,145 & 9.11 \\
            cheatsheet.com & Online-only & Culture and Technology & 2,801 & 84.33 \\
            nbcnews.com & Traditional (TV) & General & 2,602 & 9.42 \\
            cbsnews.com & Traditional (TV) & General & 2,396 & 14.28 \\
            washingtonpost.com & Traditional (Print) & General & 1,976 & 11.83 \\
                npr.org & Traditional (Radio) & General & 1,956 & 13.08 \\
                espn.com & Traditional (TV) & Sports & 1,828 & 2.33 \\
                vox.com & Online-only & General & 1,812 & 19.98 \\
            theverge.com & Online-only & Culture and Technology & 1,785 & 9.69 \\
            comicbook.com & Online-only & Culture and Technology & 1,772 & 7.35 \\
                nypost.com & Traditional (Print) & General & 1,662 & 9.51 \\
                cnet.com & Online-only & Culture and Technology & 1,608 & 13.39 \\
            washingtonexaminer.com & Traditional (Print) & General & 1,589 & 6.41 \\
            huffpost.com & Online-only & General & 1,452 & 11.71 \\
            deadline.com & Online-only & Culture and Technology & 1,410 & 3.30 \\
            engadget.com & Online-only & Culture and Technology & 1,263 & 6.34 \\
                bbc.com & Traditional (TV) & General & 1,232 & 10.92 \\
                nesn.com & Traditional (TV) & Sports & 1,181 & 2.35 \\
            gamespot.com & Online-only & Culture and Technology & 1,140 & 22.25 \\ \hline
        \end{tabular}}
    \end{center}
    \label{tab:top_embedding_outlets}
\end{table}

Articles containing embedded tweets were collected from a total of 2,089 news outlets over the collection period. On average, 27\% of the articles published by these outlets included at least one tweet. The average number of articles with embeds collected from an outlet was 66, while 1\% of the outlets (24 outlets, to be exact) published more than 1,000 articles containing embedded tweets. For outlet-level analysis, only the 361 outlets that published at least 50 articles with embedded tweets in the collection period were considered. Based on the observed data, this cut-point could indicate a substantial threshold that distinguishes outlets that use and did not use tweets as a source. 

A Pearson correlation coefficient was computed to assess if any linear relationship exists between the total articles published, $N_{total}$, and the total articles with embedded tweets published, $N_a$. This analysis presented a statistically significant positive correlation between the two variables, $r(359) = .83$, $p < .001$. Thus, the more prolific the outlet, the more articles with embedded tweets they produce. To assess the nature of the articles per news source, the number of articles published with embedded tweets was used as a measure of outlet prolificity (i.e., their story production volume) due to this linear relationship. Outlets with $500 \leq N_a \leq 2,000$ were considered ``medium'' volume outlets and those with $N_a \geq 2,000$ were considered ``large'' volume outlets. These categories were henceforth utilized for investigating the impact of outlet prolificity on embedding speed.

For each embedded tweet in the dataset, the outlet that embedded it earliest in an article was identified. Using this information, for each outlet, the fraction of first-time embeds was calculated as:
\begin{equation}
f_{1st} = \frac{N_{1st}}{N_e}
\label{eq:ffirst}
\end{equation}
where $N_{1st}$ was the total number of tweets embedded first by the outlet and $N_e$ was the total number of embedded tweets by the outlet.

To address the mechanics of tweet use in stories, at least two modes for the inclusion of tweets in articles were identified: (1) the use of a single tweet as a source, and (2) the use of multiple tweets as sources. To conceptualize these mechanics, articles with embeds were divided into two categories: \emph{single-tweet articles}, which embed only one tweet, and \emph{multi-tweet articles}, which embed multiple tweets. We found that in our sample, $61\%$ of all articles with tweets were single-tweet articles.  

Combining these two aspects of tweet embedding, then, helps answer the question of whether the outlets that are more likely to publish one-tweet stories are also more likely to publish those stories faster than other outlets. The fraction of single-tweet articles by outlet was thus computed as:
\begin{equation}
f_s = \frac{N_s}{N_a}
\label{eq:fsingle}
\end{equation}
where $N_s$ was the total number of single-tweet articles produced by the outlet.

The top outlets by embedded article volume ($N_a \geq 1,000$) were next examined to investigate embedding speed variation by outlet type, with the goal of developing a typology based on embedding speed. Results of this examination are presented in table~\ref{tab:top_embedding_outlets}.

Two typologies were developed and embed delay distributions were compared between types:

\begin{enumerate}
    \item \emph{Types based on distribution channels}: outlets were divided into two groups based on the channels on which they distribute content, online-first outlets, and traditional outlets with presence in print, television, and radio. One of the authors manually verified the platforms each outlets maintained a presence on (print, radio, television, and internet).
    \item \emph{Types based on areas of content focus}: outlets were divided into three groups based on their focus areas, namely sports-focused, culture and technology-focused, and generalized. One of the authors manually examined outlet websites and annotated each outlet with one of the three content focus types.
\end{enumerate}

\begin{figure}[h!]
    \centering
	\includegraphics[width=0.6\linewidth]{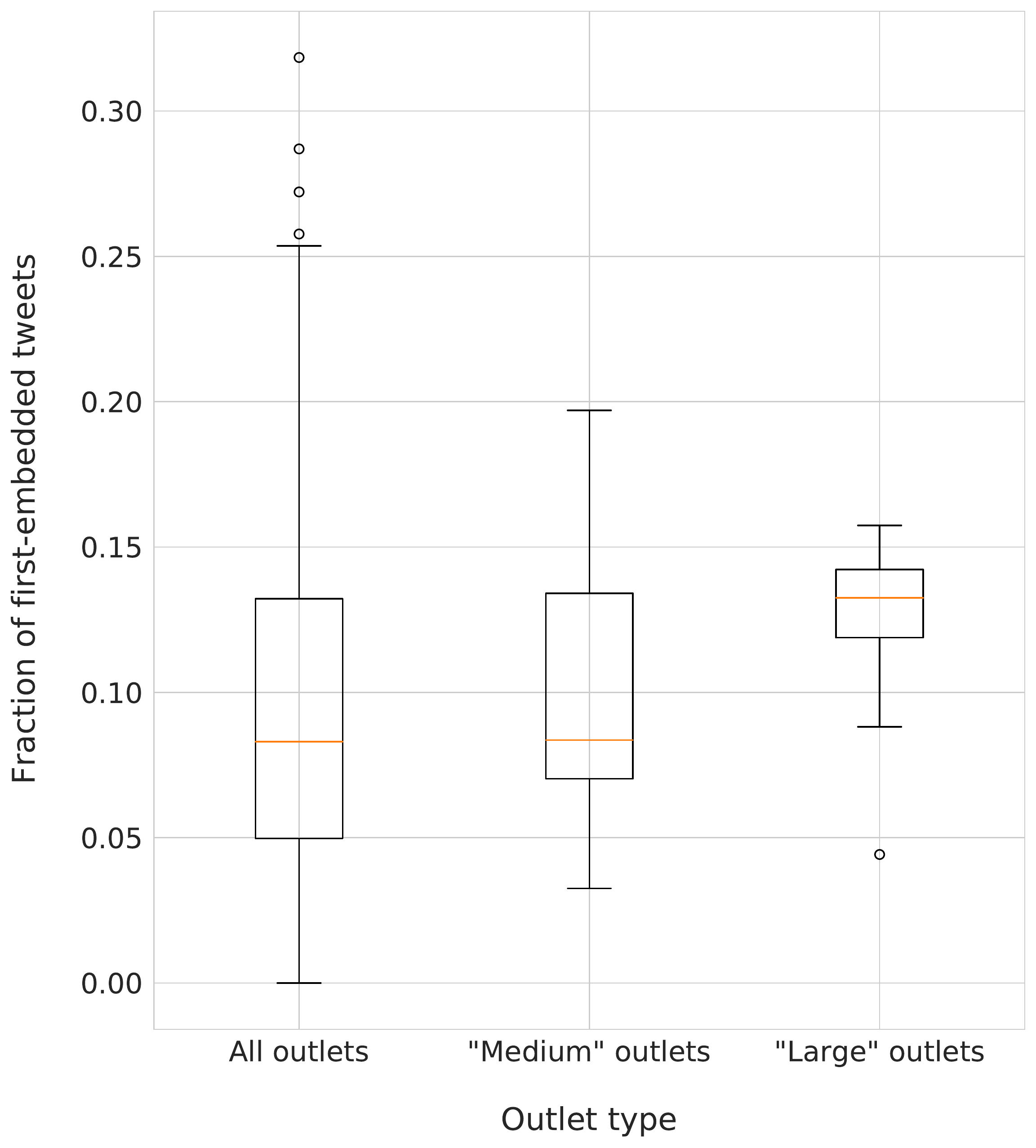}
	\caption{Distribution of first-time embed fractions by outlet.}
	\label{fig:first_embed_frac_boxplot}
\end{figure}

The next analyses sought to understand if and how tweets from different groups of users are embedded with different chronemics, and for this, a set of prominent users were examined. This set of prominent users was determined as follows. The distribution of total embeds by user was observed to be heavy-tailed, indicating that most users are only ever embedded just a handful of times. To study users in this set, the \emph{presence} of each was defined and measured by the number of days when at least one tweet from their timeline was embedded in an article from the collection period of the dataset. A typology of the top 50 users by presence was then developed with the goal of elucidating if different types of users are embedded at different speeds. Here, users were assigned the type of either ``Personality'', ``Journalist'', ``Media Outlet'', or ``Organization''. These types were assigned by manually examining the Twitter pages of users. If the account identified itself as belonging to a journalist, media outlet, or organization, those types were assigned. Otherwise, for personal accounts, the ``Personality'' type was assigned. Differences in embedding chronemics between these categories were examined, and are discussed below.

\section{Results}
These sections present the results of the above-described analysis, focusing on the effects of outlet prolificity and type in Secs.~\ref{sec:prolificity}--\ref{sec:otype}, as well as user type in Sec.~\ref{sec:utype}.

\begin{table}[hb]
    \small
    \caption{Top 20 outlets with high single-tweet article fractions and high first-time embed fractions.}
    \begin{center}
        \resizebox{\linewidth}{!}{
        \renewcommand{\arraystretch}{1.2}
        \begin{tabular}{|c|c|c|c|c|c|c|c|}
            \hline
            \textbf{Outlet} & \textbf{First time embeds} &  \textbf{First time embed fraction} & \textbf{Single tweet articles} &  \textbf{Single tweet article fraction} & \textbf{Articles with embeds} \\
\hline
         yahoo.com &    2,222 &                       0.16 &                 3,507 &                           0.57 &               14,113 \\
       thehill.com &             1,401 &                       0.16 &                 4,197 &                           0.72 &                8,943 \\
     cbssports.com &             3,996 &                       0.12 &                 2,323 &                           0.42 &               33,629 \\
           cnn.com &               708 &                       0.13 &                 2,672 &                           0.74 &                5,310 \\
     nbcsports.com &             1,111 &                       0.14 &                 1,764 &                           0.56 &                7,811 \\
       nbcnews.com &               476 &                       0.13 &                 2,073 &                           0.80 &                3,591 \\
       cbsnews.com &               479 &                       0.12 &                 1,641 &                           0.68 &                3,951 \\
          espn.com &               727 &                       0.19 &                 1,242 &                           0.68 &                3,927 \\
      theverge.com &               472 &                       0.20 &                 1,429 &                           0.80 &                2,396 \\
        nypost.com &               286 &                       0.14 &                 1,443 &                           0.87 &                2,077 \\
          cnet.com &               942 &                       0.19 &                   904 &                           0.56 &                5,013 \\
washingtonexaminer.com &               357 &                       0.13 &                   974 &                           0.61 &                2,821 \\
      deadline.com &               540 &                       0.18 &                   881 &                           0.62 &                3,070 \\
      engadget.com &               210 &                       0.14 &                 1,109 &                           0.88 &                1,476 \\
           bbc.com &               216 &                       0.12 &                   910 &                           0.74 &                1,797 \\
      gamespot.com &               321 &                       0.18 &                   856 &                           0.75 &                1,743 \\
       dexerto.com &               239 &                       0.14 &                   514 &                           0.56 &                1,757 \\
         radio.com &               140 &                       0.11 &                   412 &                           0.61 &                1,234 \\
       vulture.com &               199 &                       0.14 &                   414 &                           0.64 &                1,443 \\
      cbslocal.com &               131 &                       0.13 &                   408 &                           0.68 &                1,022 \\
\hline
        \end{tabular}}
    \end{center}
    \label{tab:first_time_and_single_tweet_outlets}
\end{table}

\subsection{Outlet prolificity and embedding chronemics}
\label{sec:prolificity}
Data-driven approaches were employed to deduce that larger outlets (conceptualized based on the amounts of stories posted) did, in general, ``win'' the race to embed tweets more often. Specifically, these data-driven approaches were based on the distribution of $f_{1st}$ (Eq.~\ref{eq:ffirst}), displayed in figure~\ref{fig:first_embed_frac_boxplot}. However, the presence of a number of medium outlets that embedded and published first more often ($f_{1st} \geq 0.1$) skewed the distribution of $f_{1st}$ higher. Outlets that covered technology (The Verge, CNET, Engadget, GameSpot, IGN, TechCrunch), many of them blogs, were prominent among these; demonstrating that journalists and bloggers specializing in topics of technology may be more inclined to use tweets as a source and report on emergent trends found on social media. Sports-focused outlets appeared to embed tweets the quickest, with prominently large shares observed across three outlets: CBS Sports (large), NBC Sports (large), and ESPN (medium) all had $f_{1st} > 0.1$.

\begin{table}[h!]
    \small
    \caption{Embed delays by outlet type.}
    \begin{center}
        \renewcommand{\arraystretch}{1.2}
        \begin{tabular}{|c|c|c|}
            \hline
            \textbf{Typology} & \textbf{Type} & \textbf{Median embed delay (hours)} \\ \hline
            Distribution & Online-only & 9.22 \\
             & Traditional & 3.21 \\ \hline
             & Sports & 1.70 \\
            Content focus & Culture and Technology & 18.23 \\
             & General & 8.27 \\ \hline           
        \end{tabular}
    \end{center}
    \label{tab:delays_by_outlet_type}
\end{table}

\subsection{Outlet type and embedding chronemics}
\label{sec:otype}
The fraction of single-tweet-articles, $f_s$ (Eq.~\ref{eq:fsingle}), demonstrated a distribution across outlets roughly concentrated in the range of $0.4 \leq f_s \leq 0.9$. Within this range, outlets could be divided into two categories: outlets with $f_{1st} \geq 0.1$, and outlets with $f_{1st} < 0.1$. By design, the second group of outlets are characterized as having \emph{both} a high fraction of single-tweet articles \emph{and} a high fraction of first-time embeds. These would be the ultimate ``winners'' of the embedding race, i.e., more likely to find newsworthy tweets quickly and report on them. The top 20 outlets from this group by number of embedded articles are presented in table~\ref{tab:first_time_and_single_tweet_outlets}. The three outlets with lowest $f_s$ among these are CBS Sports, Dexerto (an esports-focused outlet), and NBC Sports; illustrating that sports-themed outlets are more likely to publish more articles collecting multiple tweets. NBC News, CNN, BBC, The Hill, and CBS News are all part of this list as well: these mainstream, prominent ``hard news'' outlets, when they \emph{do} report on tweets, are often quick to embed and more likely to focus on a single tweet.

\begin{table}[h!]
    \small
    \caption{Top 50 users by presence.}
    \begin{center}
        \begin{tabular}{|c|c|c|c|c|}
            \hline
            \textbf{Username} &         \textbf{Type} & \textbf{Unique tweets embedded} & \textbf{Days embedded (presence)} & \textbf{Median embed delay (hours)} \\
     \hline
realDonaldTrump &  Personality &                  2,735 &                      567 &                        6.64 \\
       elonmusk &  Personality &                  1,094 &                      547 &                       23.28 \\
       RapSheet &   Journalist &                    871 &                      416 &                        2.44 \\
   AdamSchefter &   Journalist &                    748 &                      411 &                        3.68 \\
            NFL & Organization &                  1,386 &                      362 &                        3.86 \\
           espn & Media Outlet &                    541 &                      361 &                        7.43 \\
   SportsCenter & Media Outlet &                    495 &                      352 &                        9.28 \\
 BleacherReport & Media Outlet &                    520 &                      343 &                        9.78 \\
            WWE & Organization &                  2,930 &                      327 &                        0.92 \\
        atrupar &   Journalist &                    774 &                      314 &                       11.61 \\
            PFF & Organization &                    487 &                      295 &                       21.95 \\
            NBA & Organization &                    477 &                      271 &                        2.88 \\
            AOC &  Personality &                    360 &                      236 &                       11.76 \\
   TomPelissero &   Journalist &                    398 &                      235 &                        1.71 \\
     FieldYates &   Journalist &                    277 &                      218 &                        3.34 \\
            MLB & Organization &                    451 &                      212 &                        2.01 \\
        wojespn &   Journalist &                    604 &                      211 &                        2.13 \\
  ShamsCharania &   Journalist &                    371 &                      202 &                        1.52 \\
       steelers & Organization &                    588 &                      198 &                        3.39 \\
    UniverseIce &   Journalist &                    182 &                      196 &                       25.75 \\
  KimKardashian &  Personality &                    218 &                      191 &                       18.80 \\
    PFF\_College & Organization &                    336 &                      187 &                       37.66 \\
   thecheckdown & Media Outlet &                    321 &                      187 &                       15.17 \\
         Braves & Organization &                    375 &                      186 &                        1.46 \\
  ESPNStatsInfo & Media Outlet &                    245 &                      185 &                        5.90 \\
         Ravens & Organization &                    516 &                      174 &                        1.90 \\
   FortniteGame & Organization &                    187 &                      170 &                       10.80 \\
 jamisonhensley &   Journalist &                    231 &                      169 &                       19.14 \\
       Patriots & Organization &                    346 &                      164 &                        2.40 \\
            ufc & Organization &                    471 &                      164 &                        0.50 \\
      KingJames &  Personality &                    166 &                      163 &                       10.97 \\
        thehill & Media Outlet &                    187 &                      163 &                        3.52 \\
      TODAYshow & Media Outlet &                    203 &                      161 &                        3.72 \\
        Yankees & Organization &                    322 &                      159 &                        2.42 \\
    DrewShiller &   Journalist &                    308 &                      159 &                      257.28 \\
           Cubs & Organization &                    255 &                      157 &                        2.85 \\
       JoeBiden &  Personality &                    168 &                      157 &                        9.91 \\
        CBSNews & Media Outlet &                    193 &                      155 &                        6.71 \\
            ABC & Media Outlet &                    200 &                      155 &                        3.53 \\
        PGATOUR & Organization &                    922 &                      154 &                        1.87 \\
         SpaceX & Organization &                    158 &                      149 &                       14.45 \\
      CBSSports & Media Outlet &                    356 &                      146 &                        2.00 \\
 NFL\_DovKleiman &   Journalist &                    202 &                      146 &                       10.17 \\
  PitchingNinja &   Journalist &                    212 &                      145 &                        3.70 \\
     NBCSPhilly & Media Outlet &                    318 &                      145 &                        3.30 \\
     NYGovCuomo &  Personality &                    169 &                      141 &                        4.50 \\
            GMA & Media Outlet &                    154 &                      141 &                        4.67 \\
         Chiefs & Organization &                    343 &                      138 &                        1.30 \\
   MikeGarafolo &   Journalist &                    159 &                      137 &                        2.41 \\
    AlbertBreer &   Journalist &                    164 &                      135 &                        3.95 \\
     \hline            
    \end{tabular}
    \end{center}
    \label{tab:top_users}
\end{table}

A Mann-Whitney test indicated that the difference in embed delays between online-only and traditional outlets were statistically significant, $U(N_{online-only} = 56,875, N_{traditional} = 86,802) = 1,866,789,937.5, p < .001$. A one-way ANOVA was performed to compare the effect of outlet content focus on embed delays. The one-way ANOVA revealed that there was a statistically significant difference in embed delays between at least two types of outlets based on their content focus ($F(2, 136,836) = 134.92$, $p < .001$). The median embed delays by outlet type are presented in table~\ref{tab:delays_by_outlet_type}. Among the types, traditional outlets (when they chose to do so) generally embedded tweets more quickly than online-only outlets. Among the content focus-based types, outlets reporting primarily on sports were much quicker to embed tweets, with the median embed delay being under 2 hours from tweet publication. This demonstrated strong evidence for event-driven reporting, where the outlets try to capture public attention during or close to live events. In a notable contrast to this, culture and technology-focused outlets were much slower to embed tweets than both sports-focused and generalized outlets.

\subsection{User type and embedding chronemics}
\label{sec:utype}
The 50 users with the highest \emph{presence} are displayed in table~\ref{tab:top_users}. The topmost two users were extreme outliers in terms of presence, impact, and variety: former U.S. President Donald J. Trump (@realDonaldTrump) and SpaceX/Tesla C.E.O. Elon Musk (@elonmusk). Tweets from @realDonaldTrump, in addition to being embedded nearly every day of the collection period that the account was active (it was suspended permanently by Twitter on January 8, 2021), were embedded more than twice on average, while 12\% of the tweets they produced were embedded (the median was 0.1\%).

 A one-way ANOVA was performed to compare the effect of user type on embed delays. This test revealed that there was a statistically significant difference in embed delays between at least two types of users ($F(2, 34,635) = 163.78$, $p < .001$). Median embed delays by user type are displayed in table~\ref{tab:delays_by_user_type}. Tweets from the organizations in the top 50 accounts were embedded the quickest, while tweets from personal accounts were embedded the slowest. Since most of the organizations in this set were associated with sports, this result aligns with findings at the outlet level, where sports-focused outlets were seen to embed quicker than other types of outlets. The prevalence of journalists in this list of top-embedded accounts, particularly sports journalists and commentators, indicates the presence of an ecosystem where sports commentary, predictions, and general updates tweeted by journalists is quickly picked up by outlets---often the very outlets these journalists are associated with---and embedded and published within \textit{their own} articles.

\begin{table}[h!]
    \small
    \caption{Embed delays by user type.}
    \begin{center}
        \renewcommand{\arraystretch}{1.2}
        \begin{tabular}{|c|c|}
            \hline
            \textbf{Type} & \textbf{Median embed delay (hours)}\\ \hline
            Personality & 10.37\\
            Journalist & 4.54\\
            Media Outlet & 5.08\\
            Organization & 1.81\\ \hline           
        \end{tabular}
    \end{center}
    \label{tab:delays_by_user_type}
\end{table}

\section{Discussion}

In line with the tweet embedding as a journalistic practice, we expected that news outlets that produced larger amounts of stories will use tweets more often. While this was observed in the analysis, a group of ``medium'' outlets (conceptualized based on the amounts of stories created) who were often quicker than others to embed were also discovered. Many of these were technology-oriented publications and blogs. This indicated that in certain thematic areas, medium outlets competed at the same level with large outlets. The presence of these types of outlets illustrates a dynamic where by utilizing the relatively low-cost access to social media content (such as tweets), outlets can leverage immediacy to attract audience attention away from the more established institutions; shining a favorable light on the oft-discussed democratizing effects of the online ecosystem. Also, we found that more than half (i.e., 61 percent) of the stories that contained tweets, were single-tweet sourcing. Future studies should analyze more in depth, if single-tweet use was only one type of source or they complemented other types of sourcing. 

We found that when reporting on newsworthy social media content, outlets with traditional media presence (radio, television, or print) in addition to an online presence in fact acted faster than online-only outlets. This finding demonstrates the reconfiguration of journalistic values and incentives for publication. In the always-online attention economy, established outlets have come to prioritize immediacy and engagement from the audience. The quick embedding and publication of newsworthy social media content and stories driven by such content demonstrate the effects of this prioritization. 

Differing speeds of embedding by user types demonstrated a more complex picture of the tweet selection process. We noted how traditional ``credibility'' or newsworthy considerations were employed on Twitter by relying on institutional backing, consistent with the traditional sourcing of institutional ``noteworthiness'' argued in previous studies (\cite{berkowitz2019reporters}). The former president's tweets, due to the authority of the presidency as an institution, were often guaranteed selection. The celebrity status of Elon Musk led to a similar selection advantage. The institutional backing for tweets as a source creates new challenges in journalism. The former president's tweets were treated as newsworthy, even if not for endorsement reasons; thus, some scholars argued against institutional backing as the monolithic criterion in journalism coverage (\cite{parks2020ultimate}).  

Similarly, prominent figures were also featured in the news and intriguingly, journalists' tweets were rather a prominent source. Thus, in relation to authors of the source, the democratizing premises of Twitter were weaker. We observe that in spite of the obvious possibilities presented by Twitter for the inclusion of a wide variety of users as sources, journalistic practices centering on institutional credibility are re-inscribed in online spaces, as argued by previous studies where regardless of internet-democratizing aspirations, it favors ``elite'' voices (\cite{von2018sourcing}), even if regular users were also included in the news stories. This finding goes against the Twitter as democratizing sourcing hypothesis or egalitarian user integration for newsworthiness. However, it is consistent with the previous findings showing that when journalists did cite tweets from ordinary people, they are used to represent the vox populi (\cite{knight2012journalism}). These findings suggest that tweets functioned in news stories as yet another form of ``information subsidies'' of the digital age (\cite{gandy1982beyond}) where the internet has not only given citizens the means to publish, but also has given more diverse channels for the elite to spread their messages. 

\section{Conclusion and Future Work}

This study presents an extensive data collection and empirical analysis of the tweet embedding phenomenon focused on the temporal mechanics of embedding at scale. At the outlet level, distinct temporal patterns were observed by outlet type; similarly, at the user level, speed of embedding was also found to vary by user type.

In this study, we observed how cross-pollination of tweets between journalists---for multiple stories---is present in the news ecosystem, as a new reporting mechanism and impact of tweet embedding, warranting future detailed investigation. Our work likewise demonstrates the need for further qualitative examination of this ecosystem. While the results presented in this study were restricted to a focus on temporality, content-level dynamics of tweet embedding also demand deeper study: what types of tweets are embedded from the different types of users and by the different types of outlets? Do journalists employ additional mechanisms of gatekeeping when selecting certain types of content, while being more lenient when it comes to others? What are the barriers that potentially newsworthy users must overcome to be included as sources? Similarly, given that spatial dimensions, e.g, geospatial proximity appeared to mark tweet selection in political news coverage (\cite{oschatz2021twitter}), news category specificity has to be further explored.

There are several implications and future directions of the study.  In this study we excluded stories with the tweets that were published with "negative times." Even if our manual evaluation indicated that such instances are rather rare where articles are updated to include new tweets after publication, however, these ``update'' embedding events, where older articles are updated and embedded with new tweets, constitute a unique and interesting journalistic phenomenon that needs further investigation.  Future studies should analyze this phenomenon, in relation to the speed and filtering of the sourcing, i.e., how tweets are embedded in articles published \emph{after} tweet creation. In this study we focused on tweet citations automatically extracted through tags; future studies should analyze and compare different styles of tweet citation in the news stories. We also found that some news stories used multiple tweets for a story, which resembles traditional multi-source journalism practice. Thus, future studies should analyze this phenomenon with more depth to analyze types of stories and types of tweets selection practice.  

Similarly, social media sourcing has broader societal implications in the current media landscape, especially for user generated content, that is plagued by threats of misinformation and disinformation (\cite{zelenkauskaite2022creating}). While efforts of detecting such threats are promising (\cite{jussila2021text}), norms of sourcing should be adjusted, as studies have shown that institutionally-trustworthy accounts might not contain information that is trustworthy for citing (\cite{parks2020ultimate}). Further studies along credibility considerations for tweet embedding should be conducted.

\newpage
\bibliographystyle{ieeetr}  
\bibliography{references}  

\end{document}